\documentstyle[12pt,my,epsfig]{article}
\begin{document}
\thispagestyle{empty}
\lir{JINR preprint E1-98-333}
\lir{Dubna 1998}
\vspace{3.5cm}
\lic{\Large \bf Stable Algorithm for Extraction of Asymmetries } 
\lic{\Large \bf from the Data on Polarized Lepton-Nucleon}
\lic{\Large \bf Scattering}
\vskip 0.5cm
\vspace{1.cm}
\lic{N.Gagunashvili\footnote{ e-mail: gagunash@sunse.jinr.ru}}
\lic{\it Joint Institute for Nuclear Research, Dubna, Russia}
\thispagestyle{empty}
\begin {abstract}
 A new algorithm for extraction of asymmetries from
polarized lepton-nucleon scattering data is proposed.
The algorithm is stable to  set-up acceptance and/or luminosity monitor
acceptance variations. 
A statistical test for checking the data quality  is proposed.
\end{abstract}
\vspace* {1 cm}
key words: method, polarization, nucleon, asymmetry, spin \newline

\vspace* {1 cm}
\noindent
Dr.N.Gagunashvili,\newline
Laboratory of  Particle Physics,\newline
Joint Institute for Nuclear Research,\newline
Joliot-Curie str.6,\newline
Dubna,\newline
Moscow region,\newline
RU-141980, Russia.\newline
e-mail: gagunash@sunse.jinr.ru,\newline
Tel. +7 096 21 64641\newline
Fax. +7 096 21 65767

\newpage
\setcounter{page}{1}
\vskip 0.5cm

\section{Introduction} \Z
The measurement of asymmetry in deep inelastic scattering of
 polarized leptons on polarized
nucleons gives the main information for studying the 
internal spin structure of nucleons.
This approach allows to avoid the problems of normalizations
and, to some extend, acceptance corrections  to obtain then 
polarized structure functions.

Usually asymmetry is evaluated according to the formula:
\begin{equation}
A=\frac{1}{P^t P^b}\quad \frac{N^+-N^-}{N^++N^-} \quad ,
\end{equation}
where $N^+(N^-)$ is the number of scattered leptons for the 
target spin parallel (anti-parallel) to the beam spin orientation,
 $P^t$ is a value of target polarization and
$P^b$ is a value of beam polarization.
This formula is valid for the case when luminosity  and the set-up acceptance for the parallel and
anti-parallel spin orientation is the same \cite{SMCnew}. 
Usually this condition takes place when two targets with the opposite spin orientation are used for  measurements .

A more general formula is used in \cite{HERM} where measurements are carried
   out  with
only one target, 
sequentially for each target spin state:
\begin{equation}
A=\frac{N^-L^+-N^+L^-}{N^{-}L^+_P+N^{+}L^-_P}\quad ,
\end{equation}
where $L^+(L^-)$ is luminosity for each target spin state, and
$L^+_P(L^-_P)$ is luminosity weighted by the product of the beam and target
polarization values. The formula (2) is valid if the set-up acceptance
is the same and the luminosity monitor acceptance is the same
for all states of the target spin.

Another method based on  exact formulae for differences of cross-sections 
for parallel and antiparallel spin orientations is developed in \cite{nash}.
According to this approach the number of events $N_{ij}$ collected  for any
kinematical bin and for a given pair (bin) of the beam $(i)$ and target $(j)$
polarizations, can be written as
\begin {equation}
N_{ij}=\sigma_u L_{ij}+\sigma_p L_{ij}P^t_{ij}P^b_{ij} \quad , 
\end{equation} 
where $L_{ij}$ is the luminosity,
$P_i^b$, $(P_j^t)$ is the measured beam(target) polarization, 
 $\sigma_u$ is the  unpolarized part of the cross section and $\sigma_p$ is
the polarized part of the cross section. Acceptance, dead time and
other corrections are assumed to be known and properly accounted for.
The unpolarized part of cross section $\sigma_u$ and the polarized part of cross section
$\sigma_p$ are extracted from the relationship (3) by using the
likelihood method.
Asymmetry is calculated as:
\begin{equation}
A=\frac{\sigma_p}{\sigma_u}\quad .
\end{equation}

In practical cases the set-up and luminosity monitor  acceptances are not known
and they vary with time. It can be the cause of the asymmetry bias
if the methods \cite{HERM}, \cite{nash} ase used.

In this paper, we propose  an algorithm
for asymmetry extractions that is stable to acceptances  variations.
In framework of this algorithm a statistical test of the data quality check is
also developed.
The usage of the algorithm does not assume that acceptances are known or
acceptances does not vary with time when measurements are performed.

The paper consists of 4 sections. In Sect. 2 we describe an algorithm 
for asymmetry extraction. An example of using  the algorithm is described in Sect.3.  Conclusions are given in Sect. 4.

\section{Algorithm for an asymmetry extraction}\Z

Usually  the data to determine asymmetry in lepton-nucleon interactions
are collected during long periods of time , it can be done within several years.
 The beam  and target polarization,  
beam intensity, luminosity, spectrometer
acceptance, luminosity monitor acceptance, and other experimental 
conditions vary in these periods of time. To minimize the systematic  errors associated with these variations, 
 the data are divided in samples with relatively stable experimental conditions
and the target polarization is frequently reversed.

The  number
 of events $N$ collected for any kinematical bin for each sample can be written as:
\begin{equation}
N=\frac{\sigma_uCl}{c}+A\frac{\sigma_uCl}{c}P^tP^b,
\end{equation}
where $C$ is the set-up acceptance, $c$ is the luminosity monitor acceptance,
 $l$ is the measured luminosity and $P^t(P^b)$ target (beam) polarizations.
The measured luminosity can be presented as:
\begin{equation}
l=Lc \quad ,
\end{equation}
here $L$ is an unknown true value of the luminosity.
For simplicity we have assumed that the 
dead time corrections are known and properly accounted for.
We have also assumed that the target is mononuclear. 

Let us  define a
supersample as a subset of samples following one by one in time.
 We also assume that for this subset the time dependence of the variable:
\begin{equation}
s \equiv \frac{\sigma_uC}{c}\quad,
\end{equation}
can be approximated by a line
\begin {equation}
s(t)=a+bt
\end{equation}
where, $a$ and $b$, are free parameters.

Assume that the set of samples can be divided in $n$ supersamples and each $i$-th
supersample, $i=1,...,n$ contains $m_i$ samples.
The number of events $N_{ij}$ collected for any kinematical bin and 
$j$-th, $j$=1,...,$m_i$, sample of  the $i$-th supersample,  can be written as: 
\begin{equation}
N_{ij}=(a_i+b_it_{ij})l_{ij}+A(a_i+b_it_{ij})l_{ij}P^t_{ij}P^b_{ij}\quad .
\end{equation}
Our aim is to extract asymmetry from (9)
taking into account statistical errors of $N_{ij}$, luminosity, and polarizations.
The solution of this problem is found   in the framework of the
maximum likelihood method.
The task of the maximum lakelihood procedure is to find the  asymmetry, true values of
luminosities \av{l_{ij}},  target polarizations \av{P_{ij}^t}, beam polarizations \av{P_{ij}^b} and parameters $a_i$, $b_i$.

If fluctuations of polarizations and luminosity are assumed to
 be Gaussian\footnote {There is a tendency for errors that occur in many
real situations to be normally distributed due to the Central Limit Theorem. If an 
error is a sum of errors from several sources, then no matter what the probability
distribution of the separate errors may be, their sum will have a distribution
that will tend more and more to the normal distribution as the number of components increases, by the Central Limit Theorem.
 Thus, the assumption of normality is not unreasonable in most cases. In any
case we can later check the assumption by 
examining residuals.} and for $N_{ij}$ to be Poisson, the logarithm of the likelihood functional $L$, is
as follows:

\begin{eqnarray}
-ln L
=\sum_{i=1}^{n}\sum_{j=1}^{m_i}\avv{l_{ij}}(a_i+b_it_{ij})
 +A\avv{l_{ij}} \avv{P^t_{ij}} \avv{P^b_{ij}}
(a_i+b_it_{ij})-\nonumber\\
-N_{ij}ln (\avv{l_{ij}}(a_i+b_it_{ij})+A\avv{l_{ij}} \avv{P^t_{ij}} \avv{P^b_{ij}}
(a_i+b_it_{ij}))+\nonumber\\
+\frac{(l_{ij}-\avv{l_{ij}})^2}{2\delta l_{ij}^2}+
\frac{(P^t_{ij}-\avv{P^t_{ij}})^2}{2(\delta P^t_{ij})^2}
+\frac{(P^b_{ij}-\avv{P^b_{ij}})^2}{2(\delta P^b_{ij})^2}+const \quad ,
\end{eqnarray}  
where $\delta l_{ij}$, $\delta P^t_{ij}$, $\delta P^b_{ij}$ are statistical errors of $l_{ij}$,  $P^t_{ij}$,  $P^b_{ij}$ .

 Minimization of Eq. (10)
  gives asymmetry $A$,  and all other parameters 
\av{P^t_{ij}}, \av{P^b_{ij}}, \av{l_{ij}},  $a_i,b_i$.  
The complete matrix of statistical errors  is
calculated as the inverse matrix of the second derivative matrix of 
the functional (10) at its minimum \cite{bard}.

Residuals examination can be applied to the quality check of the data  used for
 asymmetry extraction \cite{smith}. 
 They are defined as:
\begin{equation}
Res \, N_{ij}=\frac{N_{ij}-N_{ij}^{fit}}{\surd N_{ij}} \quad,\\
\end{equation}
where the fitted value of the number of events 
\begin{equation}
N_{ij}^{fit} \equiv \avv{l_{ij}}(a_j+b_it_{ij})-A\avv{l_{ij}} \avv{P^t_{ij}} \avv{P^b_{ij}}(a_i+b_it_{ij})
\end{equation}
and
\begin{eqnarray}
Res \, P^t_{ij}=\frac{P^t_{ij}-\avv{P^t_{ij}}}{\delta P^t_{ij}}\qquad\qquad
\qquad \qquad \qquad \qquad \qquad \qquad \qquad \qquad \nonumber \\
Res  \, P^b_{ij}=\frac{P^b_{ij}-\avv{P^b_{ij}}}{\delta P^b_{ij}} \qquad \qquad
\qquad\qquad \qquad \qquad \qquad \qquad \qquad \qquad\nonumber \\
Res  \, l_{ij}=\frac{l_{ij}-\avv{l_{ij}}}{\delta l_{ij}}\quad.\ \ \qquad \qquad \qquad\qquad \qquad \qquad \qquad \qquad \qquad \quad 
\end{eqnarray}
The residuals plotted in the time order 
fluctuate around zero with the same variation and have no time
trend if  the quality of the data is  good and choice of supersamples is 
reasonable.
 
\section{Example of the algorithm usage}
\Z
In order  to  demonstrate the usage of the algorithm, 
pseudodata were simulated.
 We take asymmetry $A$ equal to 0.1, $\sigma_u=1.$ ,
number of samples $n$ equal to 80, luminosity $\bar L_i$, $i=1,...,80$, 
  polarizations $\bar P^t_i$,
$\bar P^b_i$, acceptances $C_i$ and $c_i$ as they are shown in Fig. 1,
 Using formulae (5) and (6), $\bar N_i$ and $\bar l_i$ are calculated. 

The measured number of events $N_i$ is simulated as Poisson distributed random variable
with a mean value equal to $\bar N_i$.
The measured values of luminosity $l_i$ and polarization $P^t_i$, $P^b_i$ were obtained by adding
fluctuations to the luminosity $\bar l_i$ and polarization $\bar P^t_i$, $\bar P^b_i$  values.
Fluctuations were simulated as Gauss distributed random variables with the mean 
value equal to zero and $\sigma$ 
equal to 2\% for luminosity, 0.5\% for target polarization and 5\% for 
beam polarization.
These pseudodata are shown in Fig. 2. 

The algorithm was used for processing the pseudodata  divided into a
different number of supersamples with the same number of samples.
The result of the algorithm application is shown in Table 1.
 The last row of Table 1 presents the asymmetry extracted
 with the algorithm from \cite {HERM}.

 In order to  calculate the average value of the asymmetry 
and   bias, the pseudodata described above were simulated 25000 times.
The average value for the asymmetry \av {A} and bias defined as the difference between the 
asymmetry average value  and its true value 0.1 are also shown in Table 1.
As follows from Table 1., the algorithm proposed in this article gives
an essentially lower bias than the algorithm in \cite {HERM}.

Residuals were analysed to choose the optimal number of supersamples.
This analysis has shown that the number of supersamples can be chosen equal to 
8 , residuals for this case are shown in  Fig. 3.  
 $N_{i}$ and its fitted value $N_i^{fit}$ are also shown in Fig. 3. \\ 
\newpage
\noindent
TABLE 1. Results of the asymmetry extraction for different numbers of supersamples. The  $\chi^2$ are calculated for the number of events $N_i$ and their fitted value $N_i^{fit}$. The last row presents the results of the algorithm \cite {HERM} application.\\

\noindent 
\begin{tabular}{|l|l|l|l|l|}
\hline
Number of  &$A$ & $\chi^2$ & $\avv{A}$&Bias \\ 
supersamples    & &  &  &  \\ \hline
\hline
1&$ 0.0799 \pm 0.1426$ &89.4987 &$ 0.1015 \pm 0.0009$&1.5\%  \\ \hline
2&$ 0.0810 \pm 0.1426$ &88.8028 &$ 0.1016 \pm 0.0009$&1.6\%   \\ \hline
4&$ 0.0679 \pm 0.1430$ &81.8742 &$ 0.0995 \pm 0.0009$&-0.5\%   \\ \hline
8&$ 0.0584 \pm 0.1442$ &63.0929&$ 0.1008 \pm 0.0009$&0.8\%   \\  \hline
16&$ 0.0676 \pm 0.1451$ &38.4059&$ 0.1007 \pm 0.0009$&0.7\%   \\  \hline
\hline
Algorithm \cite {HERM}&$0.1112 \pm 0.1456$ & & $0.1160 \pm 0.0009$&16.0\% \\ \hline
\end{tabular}
\section {Conclusions}
\Z
The algorithm for extraction asymmetry from the polarized lepton-nucleon scattering data stable to the set-up and luminosity monitor acceptances is proposed.
In framework of the developed algorithm a statistical test for checking  the data quality is 
proposed. The algorithm has no restrictions related with small
statistics of experimental events and can be applied for inclusive as well as 
for  semi-inclusive physics.
Principles of constructing the algorithm  can be used in any other field of 
 the experimental particle and nuclear physics.\\

\noindent


\noindent
{\Large \bf Acknowledgements}\\

The author would like to thank G.I.Smirnov for critical reading of the manuscript and useful discussion and  A.A.Fechtchenko for encouragement of this work.

\newpage
\begin{center}
\begin{minipage}[h]{14.7cm}
\vspace {1 cm}
\hspace*{0.8 cm}
\epsfxsize=13.7cm\epsffile{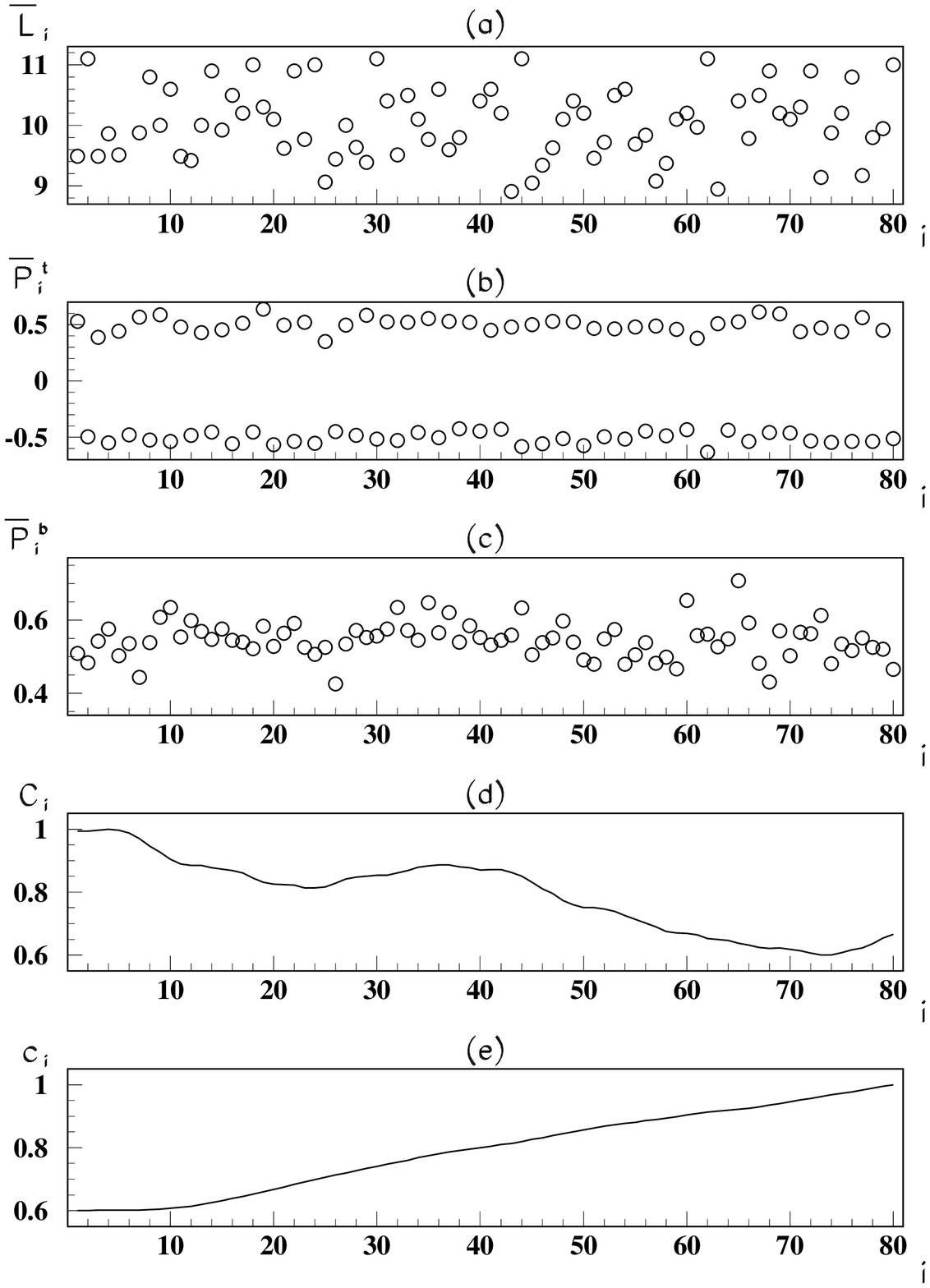} \\
Fig.1. The data for simulation  of pseudodata: (a) luminosity,
(b) target polarization, (c) beam polarization, (d) set-up acceptance,
(i) luminosity monitor acceptance.
\end{minipage}
\end{center}
\noindent
\newpage
\vspace*{1cm}
\begin{center}
\begin{minipage}[h]{14.7cm}
\hspace*{0.8 cm}
\epsfxsize=13.7cm\epsffile{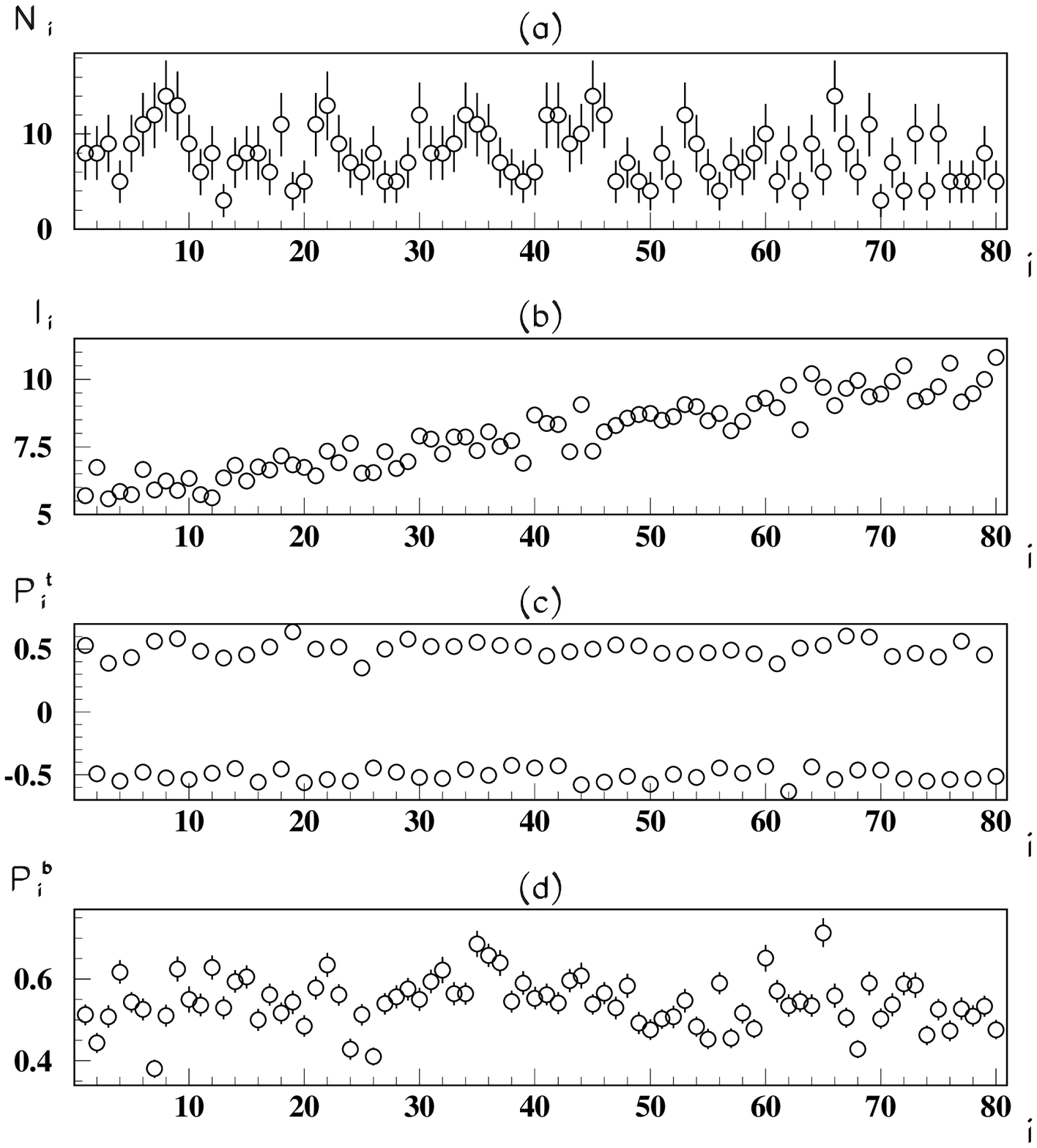}
\vspace *{-3 cm}
Fig.2. The pseudodata: (a) measured  number of events, (b) measured luminosity,
 (c) measured 
target polarization, (d) measured beam polarization.\\
\end{minipage}
\end{center}
\noindent
\vspace*{-1 cm}
\newpage
\begin{center}
\vspace * {2 cm}
\begin{minipage}[h]{14.7cm}
\hspace*{1.2 cm}
\epsfxsize=13.2cm\epsffile{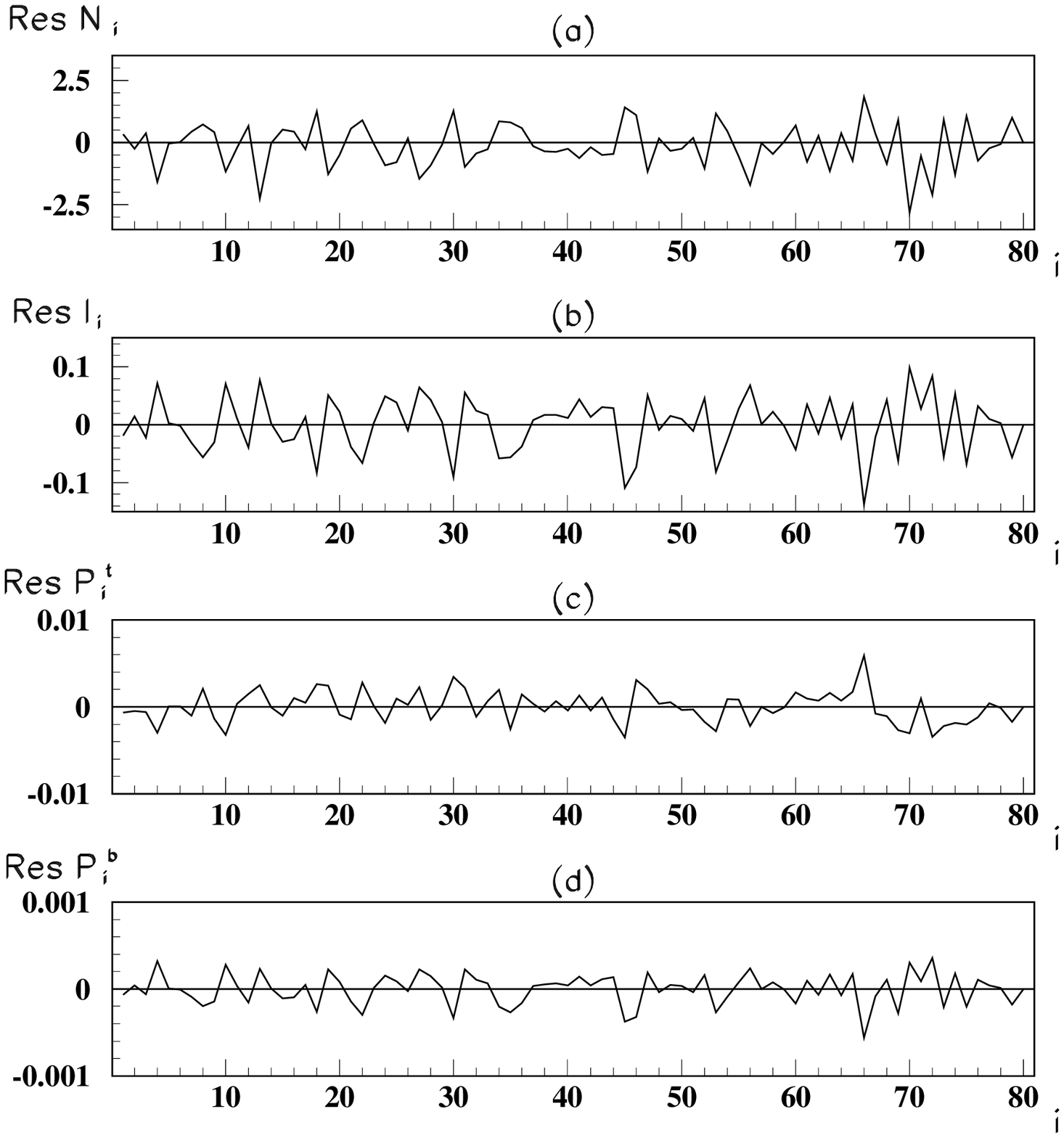}
\vspace *{-3 cm}
Fig.3. Residuals  (a) for number  of events,
 (b) for luminosity, (c)  for target \\
polarizations,
 (d) for beam polarizations.
\end{minipage}
\end{center}
\vspace*{1 cm}
\newpage
\begin{center}
\begin{minipage}[h]{14.7cm}
\hspace*{0.9 cm}
\epsfxsize=13.9cm\epsffile{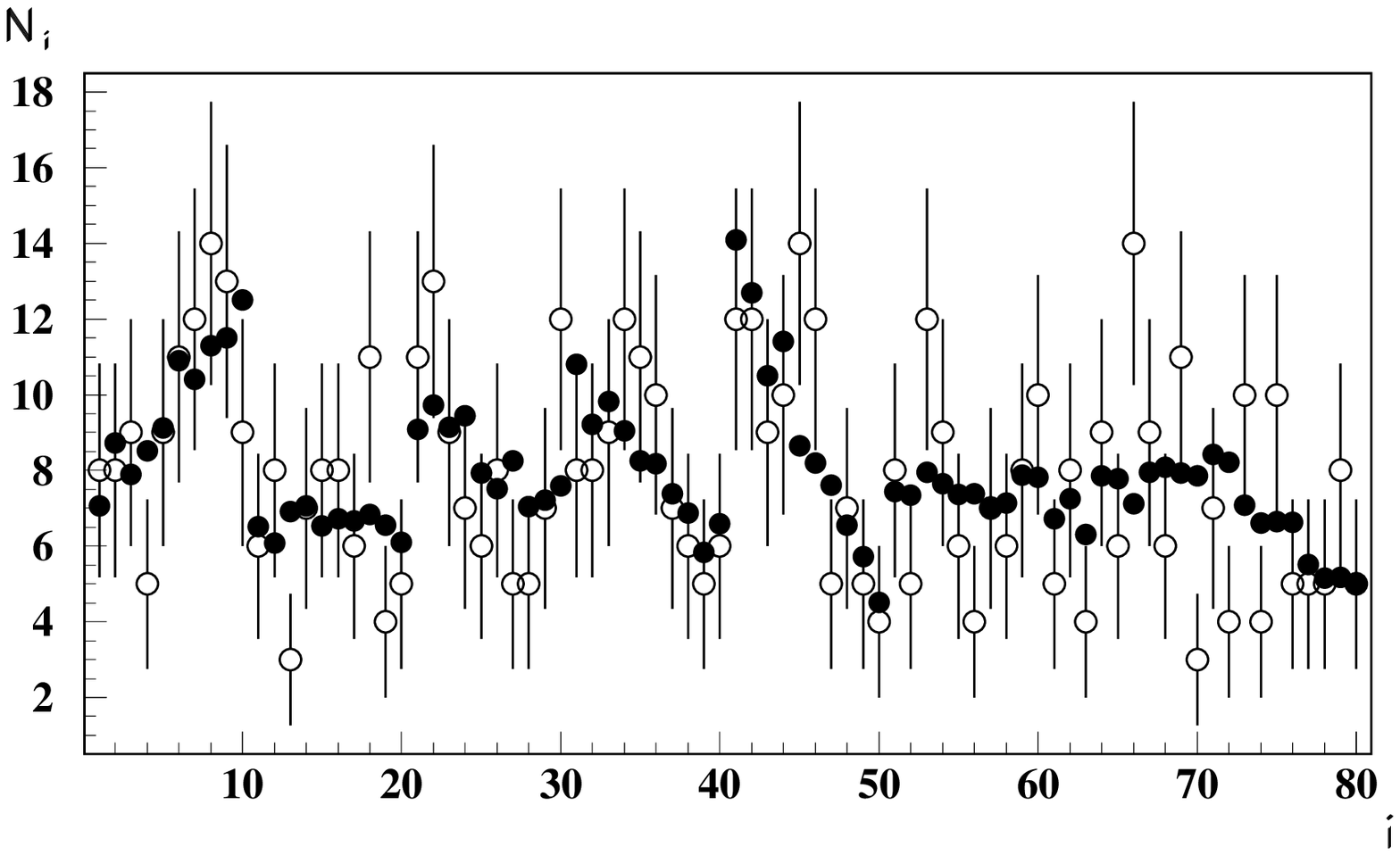}
\vspace*{-5.5 cm}
Fig.4. The number of events for  samples $N_i$ and their fitted values $N_i^{fit}$ (close\\
 cycles).\\
\end{minipage}
\end{center}

\begin {thebibliography}{99}

\bibitem{SMCnew}
  SMC, D.Adams et al., Phys.Rev. D 56 (1997) 5330. 

\bibitem {HERM}

 HERMES, K.Ackerstaff et al., Phys.Lett. B 404 (1997) 383.

\bibitem {nash}

N.D.Gagunashvili et al, Nucl.Instr. and Meth. A 412 (1998) 146.

\bibitem {bard}

Y.Bard, Nonlinear Parameter Estimation,
Academic Press New York, 1974.

\bibitem {smith}

N.R.Draper, H.Smith, Applied Regression Analysis, Johm Wiley
\& Sons, New York, 1981.

\end {thebibliography}
\end{document}